\documentclass[12pt]{article}
\usepackage{floatflt}
\usepackage{graphics}
\usepackage{rotating}
\usepackage{epsfig}
\usepackage{multicol}
\begin{document}
\title{\bf Masses and weak decay rates of doubly heavy baryons}

\author{D. Ebert$^1$, R.N. Faustov$^2$, V.O. Galkin$^2$,
  A.P. Martynenko$^3$\\
\small 
${}^1$Institut f\"ur Physik, Humboldt--Universit\"at zu Berlin,
Berlin, Germany\\
\small${}^2$Russian Academy of Sciences, Scientific Council for
Cybernetics,\\\small Moscow, Russia\\
\small${}^3$Samara State University, Samara,
  Russia} 
\date{}
\maketitle
\begin{abstract}
{Mass spectra and semileptonic decay rates of doubly heavy
  baryons are studied  in the framework of the relativistic quark
  model in the quark-diquark  approximation.}
\end{abstract}

The description of doubly heavy baryon properties acquires in the last
years the status of actual physical problem which can be studied
experimentally. The appearance of experimental data on  $B_c$ mesons,
heavy-light baryons stimulates the
investigation of heavy quark bound systems and can help in discriminating
numerous quark models. 
Recently first experimental indications of the existence of doubly charmed
baryons were published by SELEX \cite{selex}. Although these data need
further 
experimental confirmation and clarification it manifests that in the
near future the masses and decay rates of doubly heavy baryons
will be measured. This gives additional grounds
for the theoretical investigation of the doubly heavy baryon properties.
The success of
the heavy quark effective theory (HQET) in predicting 
properties of the heavy-light $q\bar Q$ mesons ($B$ and $D$)
suggests to apply these methods to heavy-light baryons, too.
The semileptonic decays of heavy hadrons present also an important tool
for determining the elements of the Cabibbo-Kobayashi-Maskawa
(CKM) matrix. 

\begin{floatingfigure}{5.23cm}
\hspace{1.3cm}\includegraphics[width=3.2cm]{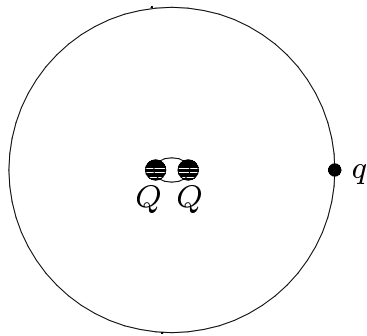}\vspace{-0.2cm}
\caption{Schematic picture of doubly heavy baryon.}\label{fig:d}
\end{floatingfigure}

Doubly heavy baryons occupy a special position among existing
baryons because they can be studied in the quark-diquark approximation
and the two-particle bound state methods can be applied. 
The two heavy quarks
compose in this case a bound diquark system in the
antitriplet colour state which serves as a localized colour
source. The light quark $q$ is orbiting around this heavy source at a
distance much larger ($\sim1/m_q$) than the source size
($\sim2/m_Q$), see Fig.~\ref{fig:d}. 
The estimates of the light quark
velocity in these 
baryons show that its value is $v/c\sim
0.7- 0.8$ and the light quark should be treated fully
relativistically.
Thus the doubly heavy baryons look effectively like a
two-body bound system and strongly resemble the heavy-light $B$ and
$D$ mesons. Then the HQET expansion in the inverse heavy
diquark mass can be performed. The ground state baryons with two heavy
quarks can be composed from a compact doubly heavy diquark of spin 0
or 1 and a light quark. According to the Pauli principle the diquarks
$(bb)$ or $(cc)$ have the spin 1 whereas diquark $(bc)$ can have both
the spin 0 and 1. 

Here we study mass spectra and semi\-leptonic decay
rates of doubly heavy baryons using the 
relativistic quark model in the quark -diquark approximation.

In the quark-diquark picture of doubly heavy baryons the bound states
of two heavy quarks and of the light 
quark and the heavy diquark are described by the
diquark wave function ($\Psi_{d}$) 
and by the baryon wave function ($\Psi_{B}$), respectively.  These
wave functions satisfy the two-particle 
quasipotential equation of the Schr\"odinger type \cite{4}
\begin{equation}
\label{quas}
{\left(\frac{b^2(M)}{2\mu_{R}}-\frac{{\bf
p}^2}{2\mu_{R}}\right)\Psi_{d,B}({\bf p})} =\int\frac{d^3 q}{(2\pi)^3}
 V({\bf p,q};M)\Psi_{d,B}({\bf q}),
\end{equation}
where the relativistic reduced mass is
\begin{equation}
\mu_{R}=\frac{E_1E_2}{E_1+E_2}=\frac{M^4-(m^2_1-m^2_2)^2}{4M^3},
\end{equation}
and the center of mass energies of particles on the
mass shell $E_1$, $E_2$ are given by
\begin{equation}
\label{ee}
E_1=\frac{M^2-m_2^2+m_1^2}{2M}, \quad E_2=\frac{M^2-m_1^2+m_2^2}{2M}.
\end{equation}
Here $M=E_1+E_2$ is the bound state mass (diquark or baryon),
$m_{1,2}$ are the masses of heavy quarks ($Q_1$ and $Q_2$) which form
the diquark or of the heavy diquark ($d$) and light quark ($q$) which form
the doubly heavy baryon ($B$), and ${\bf p}$  is their relative momentum.  
In the center of mass system the relative momentum squared on mass shell 
reads
\begin{equation}
{b^2(M) }
=\frac{[M^2-(m_1+m_2)^2][M^2-(m_1-m_2)^2]}{4M^2}.
\end{equation}

The kernel 
$V({\bf p,q};M)$ in Eq.~(\ref{quas}) is the quasipotential operator of
the quark-quark or quark-diquark interaction. It is constructed with
the help of the
off-mass-shell scattering amplitude, projected onto the positive
energy states. Here we closely follow the
similar construction of the quark-antiquark interaction in heavy mesons
which were extensively studied in our relativistic quark model
\cite{egf,pot}. For
the quark-quark interaction in a diquark we use the relation
$V_{QQ}=V_{Q\bar Q}/2$ arising under the assumption about the octet
structure of the interaction. 
The quasipotential of the quark-antiquark interaction $V_{Q\bar Q}$ 
is the sum of the usual one-gluon exchange term and the confining part
which is the mixture
of long-range vector and scalar linear potentials, where
the vector confining potential contains the Pauli term.  
The explicit expressions for $V_{QQ}$ and $V_{dq}$  and the details of
the mass spectrum calculation are given in
Ref.~\cite{r3}.  
The quark masses have the following values
$m_b=4.88$~GeV, $m_c=1.55$ GeV, $m_s=0.50$ GeV, $m_{u,d}=0.33$ GeV.  

 The masses of the
ground state  axial vector  diquarks were found to be
$M_{cc}^{AV}=3.226$~GeV, $M_{bb}^{AV}=9.778$~GeV,
$M_{bc}^{AV}=6.526$~GeV, and the mass of the scalar diquark
$M_{bc}^S=6.519$~GeV. We calculated the mass
spectra of light-quark -- heavy-diquark system first in the infinitely
heavy diquark mass limit and then 
with the inclusion of $1/M_{d}$ corrections.
In the infinitely heavy diquark mass  limit its mass and spin decouple
and the dynamics of the heavy hadron 
is determined by the light quark alone in accord with the heavy quark
symmetry. Thus the properties of heavy-light mesons and doubly heavy
baryons are similar in the limit $m_Q\to\infty$.  Inclusion of the
first order $1/m_Q$ corrections breaks the heavy quark symmetry and
leads to different splittings in mesons and baryons \cite{egf,r3}.   
Note that in our calculations we treated the light quark completely
relativistically without applying unjustified expansion in inverse
powers of its mass.

\begin{figure}
 \begin{turn}{-90} 
\epsfysize=6.4cm 
\epsfbox[110 100 518 660]{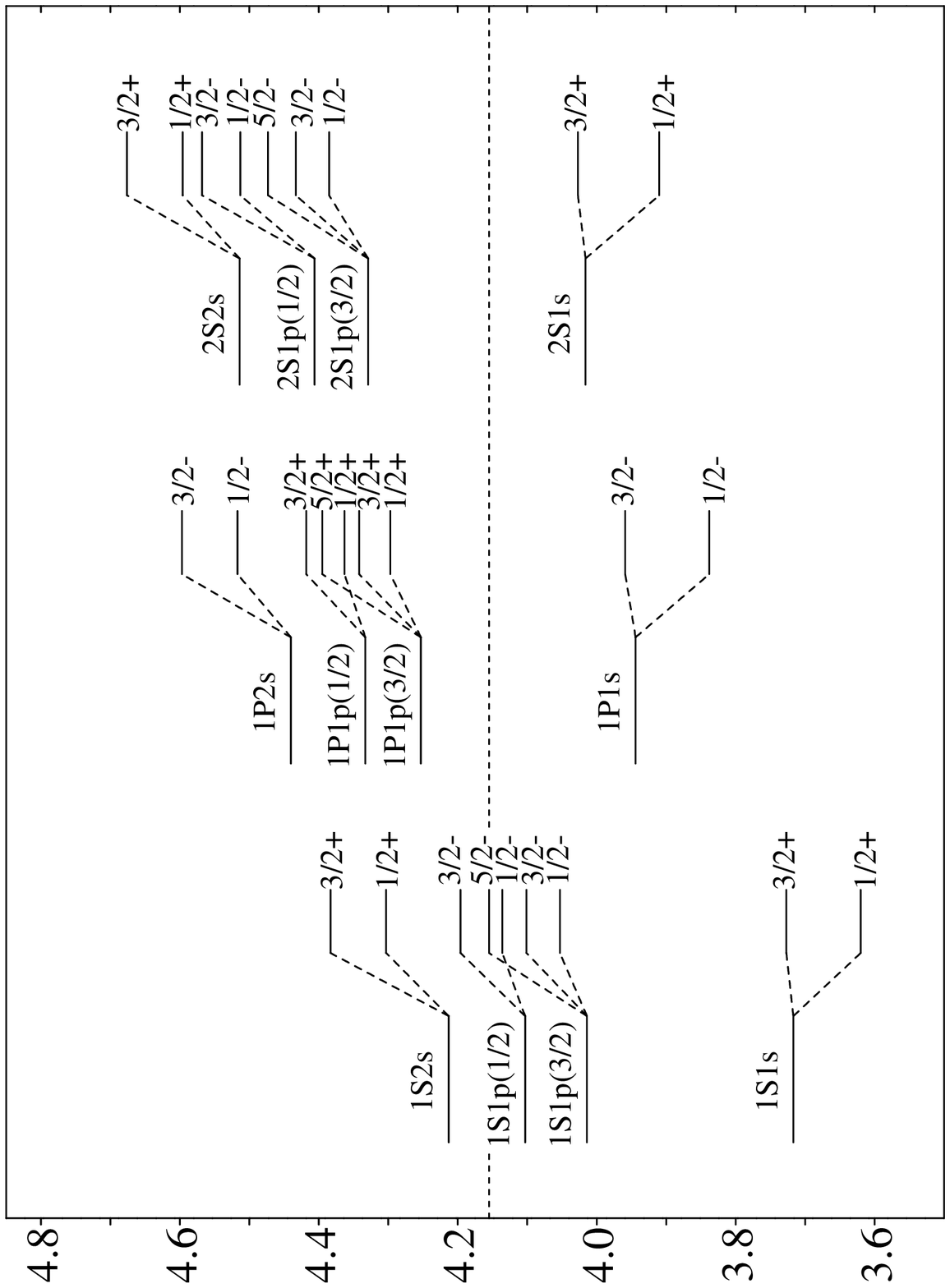}\end{turn} \ \ \ \
\begin{turn}{-90} 
\epsfysize=6.4cm 
\epsfbox[110 100 518 660]{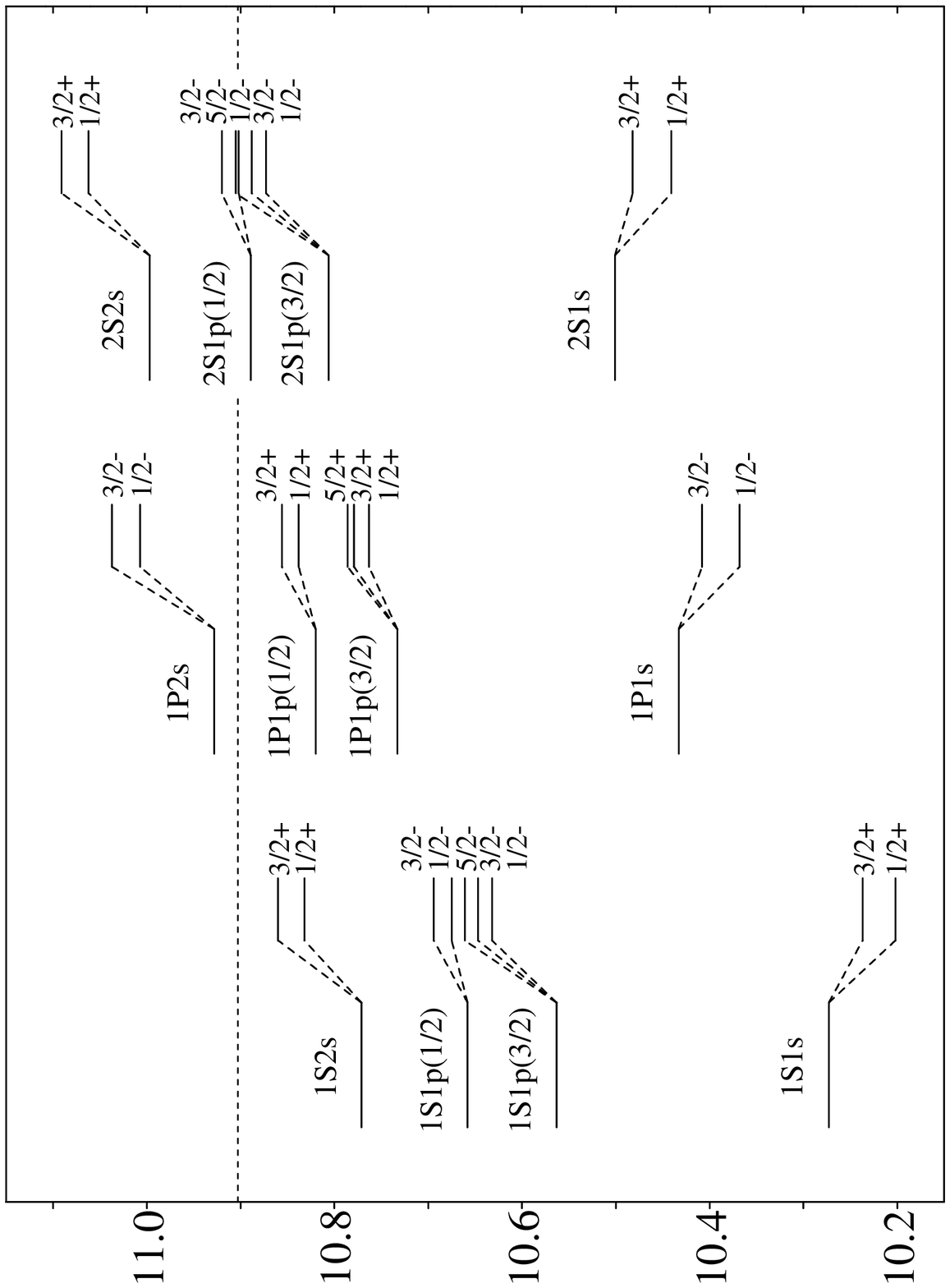}\end{turn}
\medskip
\caption{Masses of $\Xi_{cc}$ (left) and $\Xi_{bb}$ baryons (right)
  (in GeV).}\label{ms}
\vspace*{-5pt}
\end{figure}

The mass level orderings of $\Xi_{cc}$ and $\Xi_{bb}$ baryons are 
schematically shown in Fig.~\ref{ms}. There we first show our
predictions for spectra in the limit when all $1/M_{d}$
corrections are neglected. In this limit the  
$P$-wave excitations of the light quark are inverted. This means that
the mass of the state with higher light quark angular momentum $j=3/2$ is
smaller than the mass of the state with lower angular momentum $j=1/2$. 
Next we switch on $1/M_{d}$ corrections. This results in splitting of
the degenerate states and mixing of states with different $j$, which
have the same total angular momentum $J$ and parity.  The fine splitting
of $P$-levels turns out to be of the same order of magnitude as
the gap between $j=1/2$ and $j=3/2$ degenerate multiplets in the
$M_{d}\to\infty$ limit. The inclusion of
$1/M_{d}$ corrections leads also to relative shifts of the hadron
levels further decreasing this gap.  
As a result, some of the $P$-levels from different (initially
degenerate) multiplets overlap; however, the centers of levels averaged
over the heavy diquark spin remain inverted.

\begin{multicols}{2}
\hspace{-0.5cm}  \includegraphics[width=6cm]{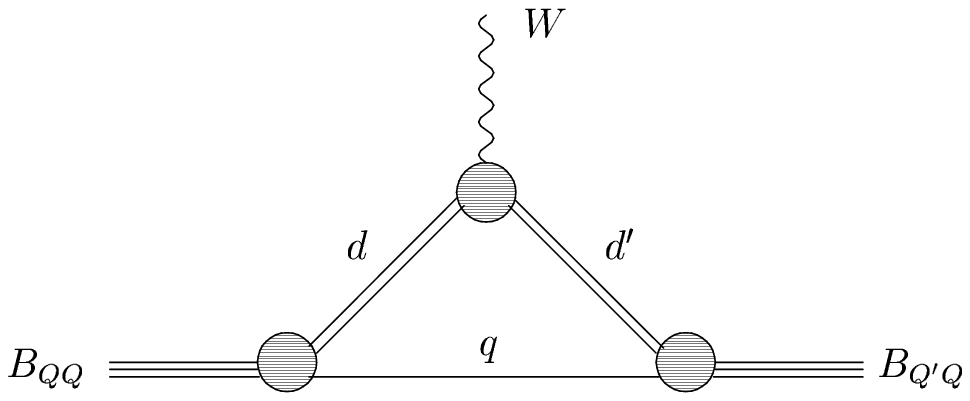}\\
  Figure 3: {Weak transition matrix element of the doubly heavy baryon
  in the quark-diquark approximation.}

\hspace{-0.5cm}  \includegraphics[width=6.5cm]{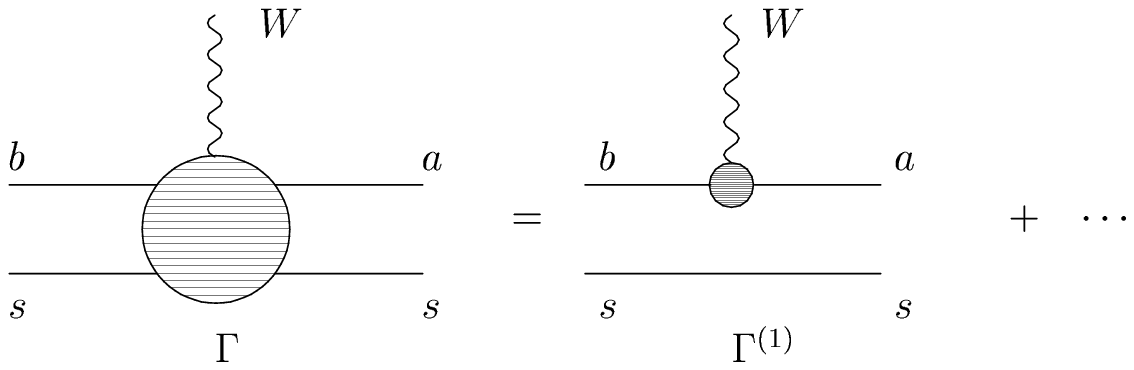}\vspace{0.3cm}\\
  Figure 4: {The leading order contribution $\Gamma^{(1)}$ to the
  diquark vertex function $\Gamma$.}
\end{multicols}

The consideration of semileptonic decays of doubly heavy baryons $(bbq)$
or $(bcq)$ to doubly heavy baryons $(bcq)$ or $(ccq)$ 
can be divided into two steps (see Fig.~3). The first step
is the study of form factors of the weak transition between initial
and final doubly heavy diquarks. The second one consists in the
inclusion of the light quark in order to compose a baryon with
spin 1/2 or 3/2.

In the relativistic quark model 
the transition matrix element between two diquark states is determined
\cite{F} 
by the convolution of the wave functions $\Psi_d$ of the initial and 
final diquarks with the two particle vertex function $\Gamma$
\begin{equation}\label{tmm}
\langle d'(Q)|J_\mu^W|d(P)\rangle=\int\frac{d^3{ p}\, d^3{
    q}}{(2\pi)^6}\bar\Psi_{d',Q}({\bf q}) 
\Gamma_\mu({\bf p},{\bf
    q})\Psi_{d,P}({\bf p}). 
\end{equation}
Here $P=vM_i$, $M_i$, $v$ denote the four-momentum, mass and
four-velocity of the initial diquark $Q_bQ_s$ and $Q=v'M_f$, $M_f$,
$v'$ denote the four-momentum, mass and four-velocity of the final
diquark $Q_aQ_s$; ${\bf p}$ and ${\bf q}$ are the relative quark-quark
momenta.   

The leading contribution to the vertex function $\Gamma_\mu$ comes
from the diagram in  Fig.~4 (index $b$ denotes the initial active  
quark, index $a$ the final active quark and index $s$ the spectator)
\begin{equation}\label{gamma1}
\Gamma_\mu({\bf p},{\bf q})
=\Gamma_\mu^{(1)}=\bar u_a({\bf q}_1)\gamma_\mu
(1-\gamma_5)u_b({\bf p}_1)\bar u_s({\bf
  q}_2)u_s({\bf p}_2) 
(2\pi)^3\delta({\bf p}_2-{\bf q}_2),
\end{equation}
where $u(p)$ is the Dirac spinor.

The transformation of the bound state wave function from the rest frame
to the moving one with four-momentum $P$ is given by  \cite{F}
\begin{equation}\label{wft}
\Psi_{d,P}({\bf p})
=D_b^{1/2}(R^W_{L_{P}})
D_s^{1/2}(R^W_{L_{P}})
\Psi_{d,0}({\bf p}),
\end{equation}
where $R^W$ is the Wigner rotation, $L_{P}$ is the Lorentz boost
from the diquark rest frame to a moving one, and  $D^{1/2}(R)$ is  
the rotation matrix.

Using this relation and the properties of the Dirac spinors and
rotation matrices we can express the matrix
element (\ref{tmm}) in the form of the trace over spinor
indices of both particles \cite{sd}. The final covariant
expression  for the transition matrix element reads
\begin{equation}\label{dd}
\frac{\langle d'(Q)|J_\mu^W|d(P)\rangle}{2\sqrt{M_iM_f}}=  
\int \frac{d^3p\, d^3 q}{(2\pi)^3}Tr\{\bar\Psi_{d'}(Q,q)\gamma_\mu(1
-\gamma_5) \Psi_d(P,p)\}\delta^3({\bf p}_2-{\bf q}_2),
\end{equation} 
where the amplitudes for the scalar ($S$) and axial vector ($AV$)
diquarks ($d$) are
given by
\begin{eqnarray}
  \label{eq:1}
 \Psi_S(P,p)\!\!\!\!&=&\!\!\!\!\!
\sqrt{\frac{\epsilon_b(p)+m_b}{2\epsilon_b(p)}}
\sqrt{\frac{\epsilon_s(p)+m_s}{2\epsilon_s(p)}}\Biggl[\frac{\hat
  v+1}{2\sqrt{2}} +\frac{\hat v-1}{2\sqrt{2}}\frac{{\tilde
    p}^2}{(\epsilon_b(p)+m_b)(\epsilon_s(p)+m_s)} \cr
&&\!\!\!\!\!-\left(\frac{\hat v+1}{2\sqrt{2}}\frac1{\epsilon_s(p)+m_s}+
\frac{\hat v-1}{2\sqrt{2}}\frac1{\epsilon_b(p)+m_b}\right)\hat{\tilde
p}\Biggr]\gamma_0\Phi_S(p),
\end{eqnarray}
\begin{eqnarray}
  \label{eq:2}
\Psi_{AV}(P,p,\varepsilon)\!\!\!\!&=&\!\!\!\!\! 
\sqrt{\frac{\epsilon_b(p)+m_b}{2\epsilon_b(p)}}
\sqrt{\frac{\epsilon_s(p)+m_s}{2\epsilon_s(p)}}\Biggl[\frac{\hat
  v+1}{2\sqrt{2}}\hat{\varepsilon}+ \frac{\hat
  v-1}{2\sqrt{2}}\cr
&&\!\!\!\!\!\times\frac{{\tilde
    p}^2}{(\epsilon_b(p)+m_b)(\epsilon_s(p)+m_s)}\hat{
  \varepsilon}   - \frac{\hat
  v-1}{2\sqrt{2}}\frac{2({\varepsilon}\cdot{\tilde 
    p})\hat{\tilde p}}{(\epsilon_b(p)+m_b)(\epsilon_s(p)+m_s)}
\cr
&&\!\!\!\!\!+\frac{\hat v+1}{2\sqrt{2}}\frac{\hat{\varepsilon}\hat{\tilde
p}}{\epsilon_s(p)+m_s} -\frac{\hat v-1}{2\sqrt{2}}\frac{\hat{\tilde p}\hat{
\varepsilon} }{\epsilon_b(p)+m_b}\Biggr]\gamma_0\gamma^5\Phi_{AV}(p).
  \end{eqnarray}
Here 
$\Phi_{d}(p)\equiv\Psi_{d,0}({\bf p})/\sqrt{2M_d}$ is the diquark
wave function in the rest frame normalized to unity and the four-vector 
$\tilde p=L_P(0,{\bf p})$.
The argument of the
  $\delta$-function in Eq.~(\ref{dd}) can  be rewritten  as
\begin{equation}
  \label{eq:df}
  {\bf p}_2-{\bf q}_2={\bf q}-{\bf
  p}-\frac{\epsilon_s(p)+\epsilon_s(q)}{w+1}({\bf v}'-{\bf v}),
\end{equation}
where $w=(v\cdot v')$.  The spectator quark
contribution factors out in all decay matrix elements. They have a
common factor  
\begin{eqnarray}
  \label{eq:sqd}
  \sqrt{\frac{\epsilon_s(p)+m_s}{2\epsilon_s(p)}}
\sqrt{\frac{\epsilon_s(q)+m_s}{2\epsilon_s(q)}}
  \!\!\!&\Biggl[&\!\!\!\!1-\sqrt{\frac{w-1}{w+1}}\left(
\frac{\sqrt{{\bf p}^2}}{\epsilon_s(p)+m_s}+\frac{\sqrt{{\bf
  q}^2}}{\epsilon_s(q)+m_s}\right)\cr
&&\!\!\!\!\!\!\!\!\!\!\!\! +\frac{\sqrt{{\bf p}^2}
\sqrt{{\bf q}^2}}{[\epsilon_s(q)+m_s][\epsilon_s(p)+m_s]}\Biggr]
= \sqrt{\frac2{w+1}}I_s(p,q).\ \ \ \ \
\end{eqnarray}
If the $\delta$-function is used to express
${\bf q}$ through  ${\bf p}$ or ${\bf p}$ through ${\bf q}$ then
$I_s(p,q)={\cal I}_s(p)$ or $I_s(p,q)={\cal I}_s(q)$ with
\begin{eqnarray*}\label{ipq}
&&\!\!\!\!\!\!\!\!\!\!{\cal I}_s(p)=
\sqrt{\frac{w\epsilon_s(p)-\sqrt{w^2-1}\sqrt{{\bf p}^2}}{\epsilon_s(p)}}
\ \theta\!\left(\sqrt{\epsilon_s(p)-m_s}-\sqrt{\frac{w-1}{w+1}}
\sqrt{\epsilon_s(p)+m_s}\right)\cr
&&\!\!\!\!\!\!\!\!\!\!+\frac{m_s}{\sqrt{\epsilon_s(p)
[w\epsilon_s(p)-\sqrt{w^2-1}\sqrt{{\bf p}^2}]}}
\ \theta\!\left(\sqrt{\frac{w-1}{w+1}}\sqrt{\epsilon_s(p)+m_s}
-\sqrt{\epsilon_s(p)-m_s}\right).
\end{eqnarray*}

The  weak current matrix elements, e.g., for the scalar to axial
vector diquark transition ($bc\to cc$) have the following covariant
decomposition 
\begin{equation}
  \label{eq:mlsav}
\frac{\langle AV(v',\varepsilon')|J^V_\mu|S(v)\rangle}{\sqrt{M_{AV}M_S}}
=ih_V(w)\epsilon_{\mu\alpha\beta\gamma} \varepsilon'^{*\alpha} v'^\beta
 v^\gamma,
\end{equation}
\begin{equation}
  \label{eq:mlsaa}
 \frac{ \langle AV(v',\varepsilon')|J^A_\mu|S(v)\rangle}{\sqrt{M_{AV}M_S}}
=h_{A_1}(w)(w+1)\varepsilon'^*_\mu-h_{A_2}(w)(v\cdot\varepsilon'^*)v_\mu
-h_{A_3}(v\cdot\varepsilon'^*)v'_\mu.
\end{equation}
These transition form factors are expressed through the overlap
integrals of the 
diquark wave functions and are given in Ref.~\cite{sd}. 
These exact expressions for diquark form factors were obtained without
any assumptions about the 
spectator and active quark masses.  For the heavy diquark system we
can apply the $v/c$ expansion. Then in the nonrelativistic limit we
get the following expressions for the form factors  
\begin{eqnarray}
  \label{eq:savff}
h_V(w)&=&[1+(w+1)f(w)]F(w),\cr
h_{A_1}(w)&=&h_{A_3}(w)=[1+(w-1)f(w)]F(w),\cr
h_{A_2}(w)&=&-2f(w)F(w),
\end{eqnarray}
where
\begin{eqnarray}
  \label{eq:ff}
  F(w)\!\!\!&=&\!\!\!\sqrt{\frac{1}{w(w+1)}}\left(1+
\frac{m_a}{\sqrt{m^2_a+(w^2-1)m_s^2}}\right)^{1/2} \cr
&&\times \int\frac{{\rm
  d}^3p}{(2\pi)^3}\bar\Phi_F\left({\bf p}+\frac{2m_s}{w+1}({\bf v}'
-{\bf v})\right)\Phi_I({\bf p})
\end{eqnarray}
and
\begin{equation}
  \label{eq:fw}
  f(w)=\frac{m_s}{\sqrt{m^2_a+(w^2-1)m_s^2}+m_a}.
\end{equation}
The appearance of the terms proportional to the function $f(w)$ is the
result of the account of 
the spectator quark recoil. Their contribution is important and
distinguishes our approach from the previous considerations
\cite{WS,Lozano}. We plot the function $F(w)$ for $bb\to bc$ and
$bc\to cc$ diquark transitions in Fig.~\ref{ff}. 
 The function $F(w)$ falls off rather rapidly, especially for the
$bb\to bc$ diquark transition where the spectator quark is the $b$
quark. Such a decrease is the
consequence of the large mass of the spectator quark and the 
high recoil momentum ($q_{\rm max}\approx m_b-m_c\sim 3.33$~GeV)
transfered.
\begin{figure}\setcounter{figure}4
\includegraphics[width=6.4cm]{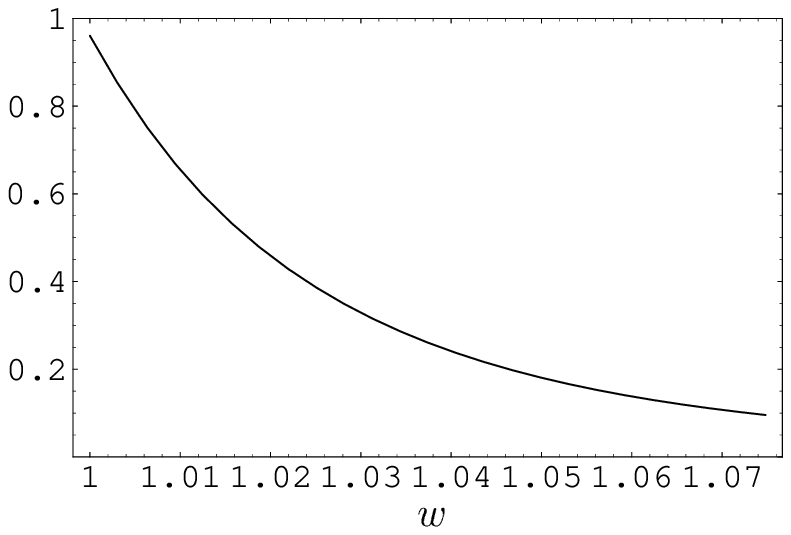}$\qquad$ 
\includegraphics[width=6.4cm]{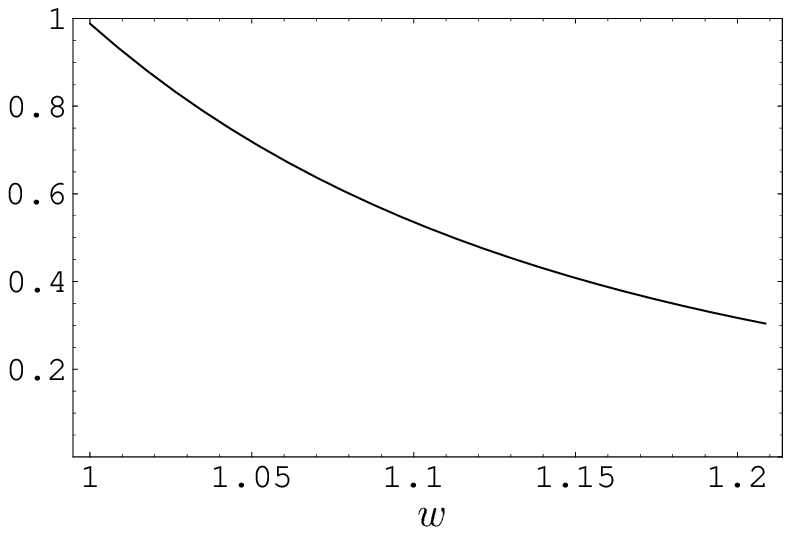}\vspace*{-10pt}
\caption{The function $F(w)$ for the $bb\to bc$ (left) and $bc\to cc$
  (right) quark transitions.} \label{ff}
\end{figure}

The second step in studying weak transitions of doubly
heavy baryons is the inclusion of the spectator light quark in
the consideration. We carry out all further calculations in the limit
of an
infinitely heavy diquark, $M_d\to\infty$, treating the light quark
relativistically.  The transition matrix element between doubly heavy
baryon states in the quark-diquark approximation (see
Figs.~3 and 4) is given by [cf. Eqs.~(\ref{tmm}) and
(\ref{gamma1})] 
\begin{equation}\label{dd1}
\frac{\langle B'(Q)|J_\mu^W|B(P)\rangle}{2\sqrt{M_IM_F}}=  
\int \frac{d^3p\, d^3 q}{(2\pi)^3}\bar \Psi_{B',Q}({\bf q})
\langle d'(Q)|J_\mu^W|d(P)\rangle
 \Psi_{B,P}({\bf p})\delta^3({\bf p}_q-{\bf q}_q),
\end{equation} 
where $\Psi_{B,P}(p)$ is the doubly heavy baryon wave function; ${\bf
  p}$ and ${\bf q}$ are the relative quark-diquark momenta.
 The baryon ground-state wave function $\Psi_{B,P}(p)$ is a product of
the spin-independent part $\Psi_B(p)$ satisfying the related
quasipotential equation (\ref{quas}) and the spin part $U_B(v)$
\begin{equation}
  \label{eq:psib}
  \Psi_{B,P}({\bf p})=\Psi_B({\bf p})U_B(v).
\end{equation}
The baryon spin wave function $U_B$ is constructed from the Dirac spinor
$u_q(v)$ of the light spectator quark and the diquark spin wave
function. The ground state spin 1/2 baryons can 
contain either the scalar  or axial vector
diquark. The former baryon is denoted by $\Xi'_{QQ'}$ and
the latter one by $\Xi_{QQ'}$. The ground state spin 3/2 baryon can be
formed only from the axial vector diquark and is denoted by $\Xi^*_{QQ'}$.
 
The amplitude for the $\Xi'_{QQ_s}\to\Xi_{Q'Q_s}$ transition  in the
infinitely heavy 
diquark limit is given by the following expression
\begin{eqnarray}
  \label{eq:ltos}
&&\!\!\!\!\!\!\frac{\langle\Xi_{Q'Q_s}(v')|J^W_\mu|\Xi'_{QQ_s}(v)\rangle}
{2\sqrt{M_IM_F}}= 
\frac{i}{\sqrt{3}}[i h_V(w)\epsilon_{\mu\alpha\beta\gamma} v'^\beta
v^\gamma 
-g_{\mu\alpha}h_{A_1}(w+1)\cr
&&+v_\mu v_\alpha h_{A_2}(w)
+v'_\mu v_\alpha h_{A_3}(w)]\bar U_{\Xi_{Q'Q_s}}(v')\gamma_5(\gamma^\alpha
+v'^\alpha)  U_{\Xi'_{QQ_s}}(v)\eta(w),\qquad
\end{eqnarray}
where $\eta(w)$ is the heavy
diquark -- light quark Isgur-Wise function which is determined by the
dynamics of the light spectator quark $q$ 
 \begin{eqnarray}
  \label{eq:eta}
\eta(w)\!\!\!&=&\!\!\!\sqrt{\frac2{w+1}} \int\frac{d^3p\, 
d^3q}{(2\pi)^3}\bar\Psi_B({\bf q})\Psi_B({\bf
p}) I_q(p,q)\cr
&&\times
\delta^3\!\left({\bf p}-{\bf q}+\frac{\epsilon_q(p)+\epsilon_q(q)}{w+1}({\bf
  v}' -{\bf v})\right).
  \end{eqnarray}
We plot the Isgur-Wise function $\eta(w)$ in Fig.~\ref{fig:eta}. In
the nonrelativistic limit for heavy quarks the diquark form factors
$h_i(w)$  contain the common factor $F(w)\eta(w)$.   

\begin{floatingfigure}{6cm}
\hspace{-1cm}  \includegraphics[width=6cm]{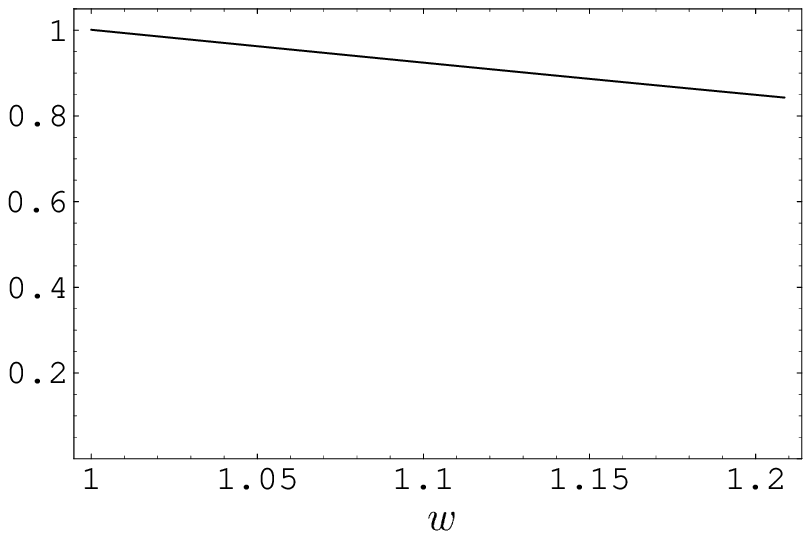}\vspace*{-10pt}
  \caption{The Isgur-Wise function $\eta(w)$ of the light quark --
  heavy diquark bound system.}
  \label{fig:eta}
\end{floatingfigure}

Our results for the semileptonic decay rates of doubly heavy
baryons $\Xi_{bb}$ and $\Xi_{bc}$ are compared with previous
predictions in Table~\ref{dhbdr}. The
results of different approaches differ substantially. Most of previous
papers \cite{Ivanov,Lozano,AO} give their predictions only for
selected decay rates. Their values agree with our in the order of magnitude.
Our predictions are smaller than the QCD sum rule
results \cite{AO} by a factor of $\sim2$.  This can be a result of our
treatment of the heavy spectator quark recoil in the heavy diquark. On
the other hand the authors of Ref.~\cite{Guo} using for
calculations the Bethe-Salpeter equation give more decay
channels. Their results are substantially higher than ours, for some
decays the difference reaches almost two orders of magnitude which
seems quite strange. E.g., for the
sum of the semileptonic decays $\Xi_{bb}\to\Xi_{bc}^{(\prime,*)}$
Ref.~\cite{Guo} predicts $\sim 6\times 10^{-13}$~GeV which almost saturates
the estimate of the total decay rate $\Gamma_{\Xi_{bb}}^{\rm
  total}\sim (8.3\pm 
0.3)\times 10^{-13}$~GeV \cite{UFN} and thus is unlikely.    

\begin{table}
\caption{\label{dhbdr}Semileptonic decay rates of doubly heavy baryons
  $\Xi_{bb}$ and $\Xi_{bc}$ 
(in $\times 10^{-14}$ GeV).}
\centerline{\begin{tabular}{cccccc}\hline
Decay & our &Ref.\cite{Guo}&Ref.\cite{Lozano} & Ref.\cite{AO} &
Ref.\cite{Ivanov} \\  \hline
$\Xi_{bb}\to\Xi'_{bc}$ & $1.64$ & $4.28$ &  &    &   \\   
$\Xi_{bb}\to\Xi_{bc}$ & $3.26$ & $28.5$ &  & $8.99$ &    \\   
$\Xi_{bb}\to\Xi^*_{bc}$ & $1.05$ & $27.2$ &  & $2.70$ & \\
$\Xi^*_{bb}\to\Xi'_{bc}$ & $1.63$ & $8.57$ &  &  &  \\ 
$\Xi^*_{bb}\to\Xi_{bc}$ & $0.55$ & $52.0$ &   &   & \\  
$\Xi^*_{bb}\to\Xi^*_{bc}$ & $3.83$ & $12.9$ &   &   & \\   
$\Xi'_{bc}\to\Xi_{cc}$ & $1.76$ & $7.76$    &   &  & \\  
$\Xi'_{bc}\to\Xi^*_{cc}$ & $3.40$ & $28.8$    &   &  & \\
$\Xi_{bc}\to\Xi_{cc}$ & $4.59$ & $8.93$ & $4.0$ & $8.87$& $0.8$ \\
$\Xi_{bc}\to\Xi^*_{cc}$ & $1.43$ & $14.1$ &   $1.2$ & $2.66$ &     \\ 
$\Xi^*_{bc}\to\Xi_{cc}$ & $0.75$ & $27.5$ &   &  & \\     
$\Xi^*_{bc}\to\Xi^*_{cc}$ & $5.37$ & $17.2$ &   &  &  \\  
\hline
\end{tabular}}
\end{table}

This work was supported in part by the DFG grant No. Eb 139/2-2.

\end{document}